\documentclass{article}
\usepackage{amsmath}
\usepackage{amsfonts}
\usepackage{amssymb}
\usepackage{amsthm}
\usepackage{PRIMEarxiv}
\usepackage{subcaption}
\usepackage{diagxy}
\usepackage{booktabs}
\usepackage{multirow}
\usepackage{hyperref}

\usepackage{todonotes}

\newcommand{\fst}[1]{\textbf{#1}}
\newcommand{\snd}[1]{\underline{#1}}
\newcommand{\sub}[2]{{#1}\sqsubseteq{#2}}

\newcommand{\interp}[1]{\ensuremath{#1^{\mathcal{I}}}}

\newcommand{\To}{\Rightarrow}
\renewcommand{\to}{\sqsubseteq}
\newcommand{\centered}[1]{\multicolumn{1}{c}{#1}}
\newcommand{\lattice}{\ensuremath{\tilde{\mathbf{C}}}}
\newcommand{\preorder}{\ensuremath{\tilde{\mathbf{R}}}}
\newcommand{\method}{CatE}

\newtheorem{theorem}{Theorem}

\usepackage[backend=biber, citestyle=numeric-comp,
sorting=none, giveninits=true, eprint=false, texencoding=utf8,
bibencoding=utf8, url=false, doi=false, isbn=false, maxnames=3,
minnames=3]{biblatex}

\AtEveryBibitem{\clearfield{month}}
\AtEveryBibitem{\clearfield{day}}
\AtEveryBibitem{\clearfield{note}}

\title{Lattice-based $\mathcal{ALC}$ ontology embeddings with saturation
}
\author{
  Fernando Zhapa-Camacho, Robert Hoehndorf \\
  Computer, Electrical and Mathematical Sciences \& Engineering Division (CEMSE) \\
  King Abdullah University of Science and Technology  Affiliation \\
  Thuwal, Saudi Arabia\\
  \texttt{\{fernando.zhapacamacho, robert.hoehndorf\}@kaust.edu.sa} \\
}

\begin{document}
\maketitle

\begin{abstract}
  Generating vector representations (embeddings) of OWL ontologies is
  a growing task due to its applications in predicting missing facts
  and knowledge-enhanced learning in fields such as
  bioinformatics. The underlying semantics of OWL ontologies is
  expressed using Description Logics (DLs). Initial approaches to
  generate embeddings relied on constructing a graph out of
  ontologies, neglecting the semantics of the logic therein. Recent
  semantic-preserving embedding methods often target lightweight DL
  languages like $\mathcal{EL}^{++}$, ignoring more expressive
  information in ontologies. Although some approaches aim to embed
  more descriptive DLs like $\mathcal{ALC}$, those methods require the
  existence of individuals, while many real-world ontologies are
  devoid of them. We propose an ontology embedding method for the
  $\mathcal{ALC}$ DL language that considers the lattice structure of
  concept descriptions. We use connections between DL and Category
  Theory to materialize the lattice structure and embed it using an
  order-preserving embedding method. We show that our method
  outperforms state-of-the-art methods in several knowledge base
  completion tasks. This is an extended version of our previous
    work \cite{cate} where we incoporate saturation procedures that
    increase the information within the constructed lattices. We make
  our code and data available at
  \url{https://github.com/bio-ontology-research-group/catE}.

\end{abstract}


%%%%%%%%%%% The article body starts:

\section{Introduction}
Ontologies are usually developed and maintained by manual curation of
experts and therefore the knowledge therein can be inconsistent or
incomplete. Traditionally, symbolic reasoners are used to test for
consistency of the knowledge within ontologies and to infer new
statements.  However, they are designed to infer statements that are
logically entailed from the ontology or knowledge base; in some cases,
it is useful to also suggest axioms that are probably true but not
entailed, leading to the task of ``ontology completion'' or
``knowledge base completion''.

From the viewpoint of knowledge graph completion~\cite{kg_completion},
we can initially define knowledge base completion as the task of predicting
``missing'' or ``novel'' axioms in a knowledge base (or ontology).
``Novel'' may be understood temporally as axioms that are added at a
later time to a knowledge base, or, more commonly, with respect to
existing axioms in the knowledge base. However, unlike knowledge
graphs, a knowledge base (ontology) has an infinitely large deductive
closure with deductively entailed statements. Those statements can be
considered ``novel'' because they do not exist in the knowledge base
but can effectively be generated by a deductive reasoner. Therefore,
knowledge base completion can have a two-fold presentation: (1)
knowledge base completion as approximate entailment, where the
completion system first generates the deductively entailed statements,
and then, with potentially lower confidence, the system generates the
non-entailed but probable statements, and (2) the completion system
generates only non-entailed statements and, optionally, has access to
information to the deductive closure.

Transversally, knowledge base completion methods can be evaluated
based on the type of axioms to complete. We distinguish between two
sub-tasks: ``TBox completion'', when the axioms to generate are
terminological and are of the form $C \sqsubseteq D$, and ``ABox
completion'', when the axioms to generate are assertional and are of
the form $C(a)$ or $r(a, b)$.  TBox completion systems have been
proposed as supporting tools to assist or automate ontology curation
procedures~\cite{2023semrec,bertsubs} or to match concepts between
ontologies~\cite{bertsubs}.  ABox completion systems are evaluated
alongside neuro-symbolic reasoners in challenges like
SemREC~\cite{2023semrec}. Furthermore, ABox completion can be regarded
as knowledge graph completion enhanced with ontological
knowledge~\cite{Hao_2019}.

Several neuro-symbolic approaches have been developed to perform the
knowledge base completion
tasks~\cite{owl2vecstar,elembeddings,box2el,bertsubs}, and most are
based on generating embeddings that preserve some logical properties
of a knowledge base.  Methods which perform knowledge base completion
follow different strategies. One type of methods corresponds to
transforming ontology axioms into graphs. Under this approach, axioms
in a DL knowledge base are transformed into a graph and then knowledge
graph completion methods are applied~\cite{owl2vecstar}. Although this
strategy has proved to be useful, this set of methods does not capture
{\em all} information in axioms and the embedding process is usually
not invertible~\cite{graph_projections}; therefore, these methods do
not allow exact inference of axioms and are often used for
similarity-based tasks.

Another type of methods for embedding DL knowledge bases constructs an
approximate model of the knowledge base.
ELEmbeddings~\cite{elembeddings} represent concepts as $n-$dimensional
balls and roles are represented as geometric translations of
concepts. By modifying the geometric structure from balls to boxes,
methods such as BoxEL~\cite{boxel} guarantee intersectional closure of
concepts (i.e., the intersection of two boxes is a box). However,
representing roles as translations can only encode one-to-one
relations. Therefore, Box$^2$EL~\cite{box2el} represents roles as two
boxes, representing the domain and the codomain of the role,
respectively. This representation enables encoding many-to-many
relations. However, all these methods target the $\mathcal{EL}^{++}$
language, which is a lightweight language that does not support the
construction of axioms involving full negation or universal
restrictions, therefore they cannot leverage more expressive
statements in DL knowledge bases. In this regard, methods such as
FALCON~\cite{falcon}, which is a method similar to Logic Tensor
Networks~\cite{ltns}, can construct an approximate model for
$\mathcal{ALC}$ knowledge bases. FALCON represents concepts as fuzzy
sets and treats logical connectives as fuzzy
operators~\cite{vanKrieken2022}. However, FALCON requires the
existence of individuals to populate the fuzzy sets, which is a
limiting factor in cases involving knowledge bases without individuals
such as the Gene Ontology (GO). Another approach for modeling the $\mathcal{ALC}$ language
is found in \cite{ozcep2023embedding} with a theoretical analysis on
the use of axis-aligned cones to represent ontology concepts.

To overcome limitations of current ontology embedding approaches, we
propose \method{}, a lattice-preserving embedding method for the
$\mathcal{ALC}$ language. Our approach relies on the fact that the
concept descriptions in a DL knowledge base can be arranged in a
lattice structure. The lattice construction of DL concepts can be
formulated in the context of Formal Concept
Analysis~\cite{Baader_2004}, using connections between DL and Modal
Logic~\cite{Schild91,modal_logic_handbook_13,modal_logic_handbook_6}
or using connections between DL and Category
Theory~\cite{cat_alc,cat_alc_reasoning}.  We use the
category-theoretical formulation and construct a lattice out of all
concept descriptions that are sub-concepts of any concept description
in the knowledge base.  After materializing the lattice we represent
its elements as vectors in an ordered-vector space. To enforce the
ordered structure of the vector space, we use an \emph{order-embedding
  method}.  We apply \method{} and show that it can outperform
state-of-the-art methods in the different forms of the knowledge base
completion task. Additionally, arranging concept descriptions in a
lattice enables the application of (partial) procedures that can
introduce new information to the lattice in terms of new elements and
morphisms. We apply partial saturation to the lattice and show that
these procedures can improve the knowledge completion performance on
some metrics such as mean reciprocal rank.  Our contributions are the
following:
\begin{itemize}
\item We propose an embedding method for $\mathcal{ALC}$ knowledge
  bases that preserves the lattice structure of the semantics of
  concept descriptions.
\item We show that our method can perform competitively on generating
  statements in the deductive closure and generating probable statements.
\item We show that our method can perform competitively in both TBox
  and ABox completion tasks.
\item  We show that partial saturation procedures can enhance the
  embedding representation of ontology concept descriptions.
\end{itemize}

\section{Preliminaries}

\subsection{Description Logics}
A Description Logic signature
$\Sigma = (\mathbf{C}, \mathbf{R}, \mathbf{I})$ consists of a set of
concept names $\mathbf{C}$, a set of role names $\mathbf{R}$, and a
set of individual names $\mathbf{I}$. In the Description Logic
$\mathcal{ALC}$, %concept descriptions can be constructed inductively;
all concept names are concept descriptions; if $A$ and $B$ are concept
descriptions, $r$ a role name, and $a, b$ are individual names, then
$A \sqcap B$, $A \sqcup B$, $\neg A$, $\exists r. A$, and
$\forall r. A$ are concept descriptions; $A \sqsubseteq B$, $A(a)$ and
$r(a, b)$ are axioms. A set of axioms is an $\mathcal{ALC}$
theory~\cite{dl_handbook}.

An interpretation $\mathcal{I}$ of an $\mathcal{ALC}$ theory consists
of an interpretation domain $\Delta^{\mathcal{I}}$ and an
interpretation function $\cdot^{\mathcal{I}}$ such that for every
concept name $C \in \mathbf{C}$,
$\interp{C} \subseteq \interp{\Delta}$; for every individual name
$a \in \mathbf{I}$, $\interp{a} \in \interp{\Delta}$; and every role
name $r \in \mathbf{R}$,
$\interp{r} \in \interp{\Delta} \times \interp{\Delta}$; and,
inductively:
\begin{equation}
  \nonumber
  \begin{split}
    \interp{\bot} &= \emptyset \\
    \interp{\top} &= \interp{\Delta} \\
    \interp{(\neg A)} & =\interp{\Delta} \backslash \interp{A}, \\
    \interp{(C \sqcap D)} &= \interp{C} \cap \interp{D} \\
    \interp{(C \sqcup D)} &= \interp{C} \cup \interp{D}, \\
    \interp{(\exists r. C)} &=\left\{a \in \interp{\Delta}
                               \mid \exists b. ((a, b) \in \interp{r} \land b \in \interp{C})\right\}\\
    \interp{(\forall r. C)} &=\left\{a \in \interp{\Delta} \mid \forall b. ((a, b) \in \interp{r} \rightarrow b \in \interp{C})\right\}
  \end{split}
  \quad \quad
  \begin{split}
   \end{split}
\end{equation}

An interpretation $\mathcal{I}$ is a model for an axiom $\sub{C}{D}$
{iff} $\interp{C} \subseteq \interp{D}$, for an axiom $B(a)$
{iff} $\interp{a} \in \interp{B}$, and for an axiom $r(a, b)$
if and only if
$(\interp{a}, \interp{b}) \in \interp{r}$~\cite{dl_handbook}. Given an
$\mathcal{ALC}$ theory $T$, an axiom is entailed from $T$ if it is
true in all models of $T$. 

\section{Construction of the Lattice Structure}
A \emph{preorder} $(P, \leq)$ contains a set $P$ equipped with a
reflexive and transitive binary relation $\leq$. A partial order is a
preorder that is also antisymmetric. A lattice is a partially ordered
set where each two-element subset has a least upper bound and greatest
lower bound. If a lattice has a greatest element, it is denoted
$\top$, and if it has a least element it is denoted
$\bot$~\cite{lattices_and_order_ch1}.

In a $\mathcal{ALC}$ theory $\mathcal{T}$, the set $\mathbf{C}$ of
concept names can be used to create arbitrarily complex and
infinitely many concept descriptions. We consider only the concept
descriptions in the knowledge base with their sub-expressions and call
this set $\tilde{\mathbf{C}}$. We furthermore denote
$\interp{\lattice{}} = \{\interp{C} \mid C \in \lattice{}\}$.

The pair $(\interp{\lattice{}}, \subseteq)$ can form a lattice where
concept descriptions $\interp{C}, \interp{D} \in \interp{\lattice{}}$
stand in a relationship if $\interp{C} \subseteq \interp{D}$.  Within
models of $\mathcal{ALC}$ theories, the relation $\subseteq$ is
reflexive and transitive. For a pair of concepts descriptions
$\interp{A}, B \in \interp{\lattice{}}$, the least upper bound is
denoted as $\interp{(A \cup B)}$ and the greatest lower bound is
denoted using $\interp{(A \cap B)}$. Additionally, for any concept
description ${X}$ it holds
$\interp{\bot} \subseteq \interp{X} \subseteq \interp{\top}$.

\begin{figure}[t]
  \centering
  \includegraphics[height=0.17\textheight]{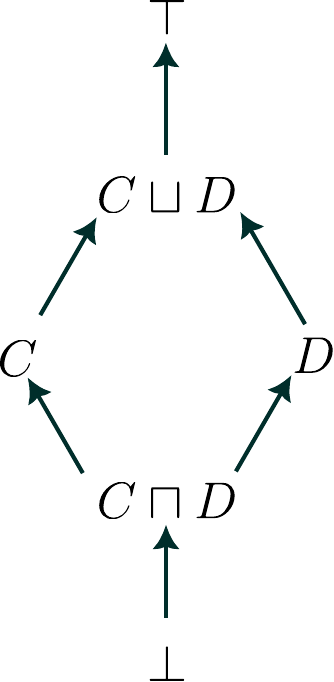}
  \caption{Lattice representation. $\bot$ is
    the bottom element and $\top$ is to top element. Arrows represent
    the $\sqsubseteq$ operator.}
  \label{fig:lattice}
\end{figure}

To represent the lattice $(\interp{\lattice{}}, \subseteq)$, we use
the syntactic representation of the axioms (where the operator is
$\sqsubseteq$ and not $\subseteq$) and denote it as
$(\lattice{}, \sqsubseteq)$ (Figure~\ref{fig:lattice}). The
representation based on $\sqsubseteq$ does not hold all the properties
of lattices; however, it is used as an intermediate structure between
the lattice $(\interp{\lattice{}}, \subseteq)$ and the embedding space
which will be introduced later (Section~\ref{sec:embedding_space}).

The concepts in $\lattice$ are materialized following a
recursive process and, depending on the type of concept descriptions,
$\lattice$ can be extended with new elements. We rely on
connections between DL and Category Theory described in \cite{cat_alc}.

\paragraph{Intersection of concepts:} Given a concept description
$A \sqcap B$ in the theory, we add the following relationships to
$(\lattice{}, \sqsubseteq)$: $A \sqcap B \to A$ and
$A \sqcap B \to B$. Additionally, for any $X$, if relationships
$X \to A \sqcap B$ are found in $(\lattice{}, \sqsubseteq)$, we add
the relationships $X \to A$ and $X \to B$
(Figure~\ref{fig:product}). Concepts $A, B$ are processed recursively.

\paragraph{Union of concepts:} Given a concept description $A \sqcup B$ in the
theory, we add the following relationships to $(\lattice{}, \sqsubseteq)$: $A \to A \sqcup B$
and $B \to A \sqcup B$. Additionally, for any $X$, if relationships
$A \sqcup B \to X$ are found in $(\lattice{}, \sqsubseteq)$, we add the relationships
$A \to X$ and $B \to X$ (Figure~\ref{fig:coproduct}). Concepts $A, B$ are processed recursively.

\paragraph{Negation of concepts:} Given a concept $\neg C$, elements
$C \sqcap \neg C$ and $C \sqcup \neg C$ are added to
$\mathbf{\tilde{C}}$. The relationships
$C \sqcap \neg C \sqsubseteq \bot$, $\top \sqsubseteq C \sqcup \neg C$
are added to $(\lattice{}, \sqsubseteq)$. Additionally, for any $X$,
if the relationship $C \sqcap X \to \bot$ is found in
$(\lattice{}, \sqsubseteq)$, we add the relationship $X \to \lnot C$,
and if the relationship $\top \to C \sqcup X$ is found in
$(\lattice{}, \sqsubseteq)$, we add the relationship $\lnot C \to X$
(Figure~\ref{fig:negation}). The concept $C$ is processed recursively.

\begin{figure}[ht]
\centering
\begin{subfigure}{.49\linewidth}
  \centering
  \begin{equation*}
    \bfig
    \Atrianglepair/>`->`>`<-`>/[X`A`A \sqcap B`B;````]
    \efig
  \end{equation*}
  \caption{Intersection}
  \label{fig:product}
\end{subfigure}
\begin{subfigure}{.49\linewidth}
  \centering
  \begin{equation*}
    \bfig
    \Vtrianglepair/>`<-`>`->`>/[A`A \sqcup B`B`X;````]
    \efig
  \end{equation*}
  \caption{Union}
  \label{fig:coproduct}
\end{subfigure}
\begin{subfigure}{.49\linewidth}
  \centering
  \begin{equation*}
     \bfig
  \morphism/@{>}@<2pt>/<0,300>[\bot`C\sqcap \neg C;]
  \morphism/@{<-}@<-2pt>/<0,300>[\bot`C\sqcap \neg C;]
  \morphism(0,300)<-250,300>[C \sqcap \neg C`C;]
  \morphism(0,300)<250,300>[C\sqcap \neg C`\neg C;]
  \morphism(-250,600)<250,300>[C`C \sqcup \neg C;]
  \morphism(250,600)<-250,300>[\neg C`C\sqcup \neg C;]
  \morphism(0,900)/@{>}@<2pt>/<0,300>[C\sqcup \neg C`\top;]
  \morphism(0,900)/@{<-}@<-2pt>/<0,300>[C \sqcup \neg C`\top;]

  \morphism/@{<-}@<2pt>/[\bot`C \sqcap X_1;]
  \morphism/@{>}@<-2pt>/[\bot`C \sqcap X_1;]
  \morphism(500,0)<0,300>[C \sqcap X_1`X_1;]
  \morphism(500,300)<-250,300>[X_1`\neg C;]

  \morphism(0,1200)/@{<-}@<2pt>/[\top`C \sqcup X_2;]
  \morphism(0,1200)/@{>}@<-2pt>/[\top`C \sqcup X_2;]

  \morphism(500,900)<0,300>[X_2`C\sqcup X_2;]
  \morphism(250,600)<250,300>[\neg C`X_2;]
  \efig
\end{equation*}
\caption{Negation}\label{fig:negation}
\end{subfigure}

   \hfill
   \caption{Lattice representations of complex concept descriptions.}\label{fig:prod_coprod}
\end{figure}

\paragraph{Existential restriction of concepts:} First, an auxiliary
preorder is constructed for DL roles, denoted as
$(\preorder{}, \sqsubseteq)$. In this preorder, elements $r, s$ stand
in a relationship $r \sqsubseteq s$ if
$\interp{r} \subseteq \interp{s}$ or if $r \sqsubseteq s$ is
entailed. \preorder{} is extended from $\mathbf{R}$ during the lattice
construction process. For any role $r$ represented in \preorder{},
elements $domain(r)$ and $codomain(r)$ are added to \lattice{}.  Given
a concept description $\exists r. C$, the relationship
$r_{\exists r. C} \to r$ is added to $(\preorder,
\sqsubseteq)$. Relationships $codomain(r_{\exists r. C}) \to C$,
$domain(r_{\exists r. C}) \to \exists r. C$ and
$ \exists r. C \to domain(r_{\exists r. C})$ are added to
$(\lattice{}, \sqsubseteq)$.  Additionally, if there are roles
$s \in \preorder$ with relationships $s \to r$ and
$codomain(r) \to C$, the relationship
$domain(s) \to domain(r_{\exists r. C})$ is added to
$(\lattice{}, \sqsubseteq)$. The concept $C$ is processed recursively.

\paragraph{Universal restriction of concepts:} Given a concept
description $\forall r. C$, the element $\lnot \exists r. \lnot C$ is
added to \lattice{} with relationships
$\forall r. C \to \lnot \exists r. \lnot C$ and
$\lnot \exists r. \lnot C \to \forall r. C$. Furthermore, if there are
roles $s \in \preorder$ with relationships $s \to r$ and
$domain(s) \to \forall r. C$, the relationship $codomain(r) \to C$ is
added to $(\lattice{}, \sqsubseteq)$. Concepts
$\lnot \exists r. \lnot C$, $\lnot C$ and $C$ are processed
recursively.

\paragraph{Subsumption axioms:} Axioms $C \sqsubseteq D$ are
incorporated directly to the lattice. Additionally, relationships
$\top \to \neg C \sqcup D$ are added to $(\lattice{},
\sqsubseteq)$. Concepts $C$ and $D$ are processed recursively.

\paragraph{Class assertion axioms:} Given an axiom $C(a)$, we
construct the element $\{a\}$ in \lattice{} with the following
relationships: $\bot \to \{a\}$, $\{a\} \to C$ and $\{a\} \to \top$.

\paragraph{Role assertion axioms:} Given an axiom $r(a, b)$, we
construct elements $\{a\}, \{b\}$ in \lattice{} with the following
relationships: $\bot \to \{a\}$, $\{a\} \to \top$, $\bot \to \{b\}$,
$\{b\} \to \top$ and $\{a\} \to \exists r. \{b\}$.  \newline

Every operator ($\sqcap \mid \sqcup \mid \neg \mid \exists \mid
\forall \mid \sqsubseteq$) introduces a constant number of elements
into \lattice{} and a constant number of relationships in
$(\lattice{}, \sqsubseteq)$. Therefore, for a formula in the knowledge
base with $n$ operators the space and time complexity to process it is $O(n)$.
The lattice construction process is not complete since we consider a
subset $\lattice$ from the infinite set $\mathbf{C}$ of
possible concept descriptions.

\subsection{Saturation procedures}
The lattice structure of concept descriptions allow for implementing
partial saturation procedures. By ``saturation'' we refer to the
process of adding new elements and morphisms to the lattice until a
fixed point is reached. However, the saturation process we perform is
partial, in the sense that the fixed point might not be actually
obtained but some additional information is added to the
lattice. Since the lattice is equipped with a transitive relation, an
inmediate saturation rule is to compute the transitive closure of the
lattice. Additionally, as specified in \cite{cat_alc_reasoning},
certain deduction rules can be transformed into partial saturation
procedures.  We specify the rules below in the form of
$precondition \To consequence$, where $precondition$ denotes the set
of morphisms existing in the lattice and $consequence$ denotes the set
of elements and morphisms to be added to the lattice. Therefore, for
elements $C, D, E\in \lattice{}$ and for elements in
$r \in \preorder{}$:

\begin{align}
  \centering
  C \to \lnot D & \To D \to \lnot C \label{eq:24}\\ %eq 24
  C \sqcap D \to \bot & \To C \to \lnot D \label{eq:25}\\ %eq 25
  \bot \to C \sqcup D & \To \lnot C \to D \label{eq:26}\\ %eq26
  C \to D, D \to E & \To C \to E \label{eq:transitivity}\\
  C \to D & \To \exists r. C \to  \exists r. D \label{eq:36} %36
\end{align}

Equations \ref{eq:24}, \ref{eq:25}, \ref{eq:26} are applicable to a subset of the
asserted morphisms in the lattice and all of them introduce one new
element to the lattice. Equation \ref{eq:transitivity} correspond to
the transitive closure of the lattice and only introduces new
morphisms but not new elements. Equation \ref{eq:36} is applicable to
all morphisms in the lattice and introduces $2 \times |\preorder|$
elements per each morphisms in the lattice. Due to the large space
complexity of Equation \ref{eq:36}, we do not consider it in our
analysis.

\subsection{Embedding into an ordered-vector space}~\label{sec:embedding_space}

With the structure $(\lattice{}, \sqsubseteq)$ in place, we proceed to
embed it into an ordered-vector space. This step is crucial for
preserving the hierarchical relationships within the lattice, ensuring
that our embeddings reflect the inherent ordering of concepts
descriptions. We use an ordered-vector space $(X, \preceq)$ over
$\mathbb{R}^n$ where, for elements in $a, b \in X$ with
$a=(a_1,..., a_n)$ and $b=(b_1,..., b_n)$, $a\preceq b$ if and only if
$a_1\leq b_1, ..., a_n\leq b_n$.
\begin{theorem}[$(X, \preceq)$ is a partial order]~\label{th:ordered_vector_space}
  The pair $(X, \preceq)$ over $\mathbb{R}^{n}$, where for elements
  $a, b \in X$ with $a=(a_1,..., a_n)$ and $b=(b_1,..., b_n)$,
  $a\preceq b$ if and only if $a_1\leq b_1, ..., a_n\leq b_n$, is a
  partial order.
\end{theorem}

\begin{proof} We demonstrate for each property of a partial order:
  \begin{enumerate}
  \item Reflexivity $(\Rightarrow)$: Let $a \in X$ with $a\preceq a$.
    By definition, we have $a_i \leq a_i$ for any $i$. $(\Leftarrow)$: Let $a\in X$. Since $a_i
    \leq a_i$ for any $i$, then $a\preceq a$.
  \item Transitivity $(\Rightarrow)$: Let $a, b,c \in X$. If
    $a\preceq b$ and $b \preceq c$, we have that $a_i \leq b_i$ and
    $b_i \leq c_i$; therefore, $a_i \leq c_i$ for any
    $i$. $(\Leftarrow)$: Let $a, b,c \in X$ with $a_i \leq b_i$ and
    $b_i \leq c_i$ for any $i$. It follows that $a_i \leq c_i$, which
    implies $a \preceq c$.
  \item Antisymmetry $(\Rightarrow)$: Let $a, b \in X$. If $a\preceq b$ and $b\preceq
    a$, it follows that $a_i \leq b_i$ and $b_i \leq a_i$. Therefore,
    $a_i = b_i$ and $a=b$. $(\Leftarrow)$: Let $a, b \in X$ with $a_i =
    b_i$ for any $i$. It implies that $a_i \leq b_i$ and $b_i \leq
    a_i$, therefore, $a\preceq b$ and $b \preceq a$.
  \end{enumerate}
\end{proof}

Consequently, we introduce a parameterized function $f_\theta$ which
maps objects in $(\lattice{}, \sqsubseteq)$ to the ordered-vector
space $(X, \preceq)$ over $\mathbb{R}^n$.  In this way, we intend
$f_\theta$ to be a lattice-preserving function of
$(\lattice{}, \sqsubseteq)$.  Since $f_\theta$ is unknown, our task is
to find the set of parameters $\theta \in \Theta$ that accommodates to
the intended structure of the space $X$. We optimize $f_\theta$ using
gradient descent. We use the following order-preserving scoring
function~\cite{ordere}:

\begin{equation}
  \label{eq:ordere}
    s(A, B) = ||\max(0,f_\theta(A)-f_\theta(B))||_2
\end{equation}
for elements $A, B\in \lattice{}$ with a relationship
$A \sqsubseteq B$. If $f_\theta(A) \preceq f_\theta(B)$, then
$s(A,B) = 0$, and otherwise $s(A,B) > 0$. We apply the following loss
function to all relationships
$A\sqsubseteq B \in (\lattice{}, \sqsubseteq)$:

\begin{equation}
  \label{eq:loss_fn}
  \mathcal{L} = \sum_{A\to B \in
    (\lattice{}, \sqsubseteq)}\sum_{A \to {B}'\notin
    (\lattice{}, \sqsubseteq)} {s(A, B) + \max{(0, \gamma - s(A, {B}'))}}
\end{equation}

Relationships $A\sqsubseteq {B}' \notin (\lattice{}, \sqsubseteq)$ are
called negative samples and are generated by replacing $B$ in an
existing relationship $A\sqsubseteq B$ by a corrupted entity ${B}'$
obtained by random sampling in a uniform distribution. The parameter
$\gamma$ is a margin parameter enforcing a minimum score value of the
negative samples.

We show that the space $X$ gets a partial order structure whenever the
loss function $\mathcal{L}=0$.
\begin{theorem}[Lattice-preserving
  embeddings]~\label{th:lattice_preserving} Let $\mathcal{O}$ be a
  $\mathcal{ALC}$ theory with signature
  $\Sigma = (\mathbf{C}, \mathbf{R}, \mathbf{I})$ and
  $(\lattice, \sqsubseteq)$ the lattice of concepts descriptions
  generated from $\mathcal{O}$. Let $(X, \preceq)$ be an
  ordered-vector space where for elements $a, b \in X$ with
  $a=(a_1,..., a_n)$ and $b=(b_1,..., b_n)$, $a\preceq b$ if and only
  if $a_1\leq b_1, ..., a_n\leq b_n$. Let $f_\theta$ be a function
  mapping objects from \lattice{} to $X$. If $\mathcal{L} = 0$, then
  $f_\theta$ is a lattice preserving function of
  $(\lattice, \sqsubseteq)$ into $(X, \preceq)$.
\end{theorem}

\begin{proof}
  Let us assume that $\mathcal{L}=0$ and there exist a relationship
  $A \sqsubseteq B$ in the lattice such that
  $f_\theta(A) \npreceq f_\theta(B)$, meaning that the order is not
  preserved in the vector space $X$. Reordering the definition of $L$
  in Equation~\ref{eq:loss_fn}, we have that
  $\mathcal{L} = s(A, B) + K$, where $K$ is a non-negative
  number. Therefore, since $\mathcal{L} = 0$, it follows that
  $s(A, B) = ||\max{0, f_\theta(A) - f_\theta(B)||} =
  0$. Consequently, we have that $f_\theta(A) \preceq f_\theta(B)$,
  which leads to a contradiction.

  Now that we have shown that any relationship $A
  \sqsubseteq B$ in the
  lattice $(\lattice{}, \sqsubseteq)$ is preserved as $f_\theta(A)
  \preceq f_\theta(B)$ in $(X, \preceq)$, we now verify that
  $f_\theta$ preserves partial-order properties:
  \begin{enumerate}
  \item Reflexivity: Let $A\sqsubseteq A$ be a relationship in
    $(\lattice{}, \sqsubseteq)$. Since $\mathcal{L} = 0$, it implies
    that $f_\theta(A) \preceq f_\theta(A)$.
  \item Transitivity: Let $A \sqsubseteq B$ and $B \sqsubseteq C$ be
    relationships in $(\lattice{}, \sqsubseteq)$. Since
    $\mathcal{L}=0$, it follows that $f_\theta(A) \preceq f_\theta(B)$
    and $f_\theta(B) \preceq f_\theta(C)$ and, by the transitive
    property of $\preceq$ (Theorem~\ref{th:ordered_vector_space}),
    $f_\theta(A) \preceq f_\theta(C)$.
  \item Antisymmetry: Let $A \sqsubseteq B$ and $B \sqsubseteq A$ be
    relationships in $(\lattice{}, \sqsubseteq)$. Since
    $\mathcal{L}=0$, it follows that $f_\theta(A) \preceq f_\theta(B)$
    and $f_\theta(B) \preceq f_\theta(A)$ and, by the antisymmetry
    property of $\preceq$ (Theorem~\ref{th:ordered_vector_space}),
    $f_\theta(A) = f_\theta(B)$.
  \end{enumerate}
\end{proof}

\section{Evaluation}

To show the effectiveness of our method, we evaluate on the following
tasks: (1) generation of entailed axioms and (2) generation of
probable axioms.  In the task of generating entailed axioms, we use
the ORE1 dataset from SemREC~\cite{2023semrec} and generate axioms of
the form $C(a)$, where $C$ is a concept name and $a$ is an individual.
In the case of generating probable axioms, we constructed datasets
using GO~\cite{go} and FoodOn~\cite{foodon} to generate axioms of the
form $C \sqsubseteq D$, where $C,D$ are concept names. For each case,
  we also show that partially saturating the constructed lattice
  impacts the performance of axiom generation. Additionally,
we applied our method to the biomedical task of predicting
protein--protein interactions. This task is a form of generation of
probable statements of the form $r(a,b)$, where $r$ is a role and
$a, b$ are individuals. We show information about datasets in
Table~\ref{tab:datasets}.

\begin{table}[t]
  \centering
  \caption[Datasets statistics]{Number of axioms in training, validation and testing ontologies and number of relationships in the corresponding training lattices.}\label{tab:datasets}
  \begin{tabular*}{\columnwidth}{@{\extracolsep\fill}lrrrr@{\extracolsep\fill}}
    \toprule
    Dataset & Training  & Validation & Testing    & Lattice \\ \midrule
    ORE1 & 61245 & 7578 & 15157 & 364849\\
    FoodOn & 34224 & 2977 & 5957 & 631423 \\
    GO & 81844 & 7260 & 14521 & 1257443\\ 
    PPI & 351435 & 12038 & 12040 & 4479085\\
    \bottomrule
  \end{tabular*}
\end{table}

\subsection{Experimental Setup}

To find the optimal hyperparameters for our method, we performed a
grid search over parameters: embedding dimension $\in [50, 100, 200]$,
margin ($\gamma$) $\in [0, 0.01, 0.1, 1]$, number of negative samples $\in [1, 2, 4]$,
batch size $\in [8192, 16384, 32768]$, and learning rate
$\in [1e^{-5}, 1e^{-4}, 1e^{-3}, 1e^{-2}]$. We used the Adam
optimizer~\cite{adam} with a Cyclic Learning Rate
scheduler~\cite{cyclic_lr}.

As baseline methods we selected those approaches that use only the
ontology axioms, without any external knowledge such as
text\cite{owl2vecstar,bertsubs}. Therefore, we selected
ELEmbeddings~\cite{elembeddings} and Box$^2$EL~\cite{box2el}. We used
the implementations provided in the mOWL library~\cite{mowl}. To
obtain optimal parameters for baseline methods, we performed a grid
search over embedding dimension $\in [50, 100, 200]$, margin $\in [0, 0.01, 0.1]$
batch size $\in [5000, 10000, 20000]$ and learning rate 
$\in [1e^{-5}, 1e^{-4}, 1e^{-3}]$. Additionally, we compared with
 FALCON~\cite{falcon}; however, due to high memory and time
requirements, we were unable to test different hyperparameters for
this method. All selected hyperparameters are provided in the
Appendix~\ref{app:hyperparameter}.

% For the protein--protein interaction prediction task, we compare
% against methods found in \cite{semantic_similarity}, \cite{boxel},
% \cite{falcon} and ~\cite{box2el}, and used their reported results.

We report a variety of rank-based metrics such as Mean Rank (MR), Mean
Reciprocal Rank (MRR), Hits@3, Hits@10, Hits@100 and ROC AUC.
% In the protein--protein interaction prediction task, we report both
% raw and filtered metrics.
In all tasks we report filtered metrics only and filter statements
from the training set. In the task of generating axioms $C(a)$, we
additionally filter statements from the deductive closure of the
training set.

\subsection{Generating Entailed Axioms $C(a)$}

The SemREC challenge~\cite{2023semrec}, which evaluates neuro-symbolic reasoners,
provides a number of benchmark datasets. We selected a representative
data set called ORE1.  We used the ORE1 dataset to test our method on
the task of predicting axioms $C(a)$, where $C$ is a concept
description and $a$ is an individual. We perform a ranking-based
evaluation, where we rank every testing statement $C(a)$ against every
$C'(a)$ where $C'$ is a named concept. We show results in
Table~\ref{tab:membership_prediction}, where we can see \method{}
performs better than baseline methods across all metrics.

\begin{table}[t]
  \centering
  \caption[Membership Prediction]{Prediction of axioms $C(a)$ where
    $C$ is a concept and $a$ is an individual. We selected the ORE1
    dataset proposed in \cite{2023semrec}.}\label{tab:membership prediction}~\label{tab:membership_prediction}
  \begin{tabular*}{\columnwidth}{@{\extracolsep\fill}lrrrrrr@{\extracolsep\fill}}
    \toprule
    Method       & {MR}      & MRR         & Hits@3           & Hits@10     & Hits@100    & AUC  \\ \midrule
    ELEmbeddings & \snd{105} & \snd{0.12} & \snd{0.08}       & \snd{0.22} & \snd{0.87} & \snd{0.99} \\
    Box$^2$EL    & 122       & 0.10 & 0.08 & 0.18 & 0.70 & 0.98 \\
    FALCON       & 603       & 0.02        & 0.00             & 0.02        &	0.34 & 0.92 \\
    \method{} & \fst{37} & \fst{0.18}& \fst{0.10} & \fst{0.51} & \fst{0.96} & \fst{0.99} \\
    \bottomrule
  \end{tabular*}
\end{table}

\subsection{Generating Probable Axioms $C\sqsubseteq D$}

To evaluate on the task of generating probable axioms, we generate two
benchmark sets following procedures designed in previous
methods~\cite{owl2vecstar,emel}. We create two datasets using the Gene
Ontology~\cite{go} and the Food Ontology~\cite{foodon}. In each
ontology we remove 30\% of the axioms $\sub{C}{D}$ uniformly at random
and distribute 10\% for validation and 20\% for testing. The training
set contains the 70\% of the subsumption axioms together with the
other axioms existing in the ontology.

We focus on the prediction of subsumption axioms $\sub{C}{D}$ and
perform a rank-based evaluation ranking scores of axioms of interest
$C\sqsubseteq D$ over all axioms $C\sqsubseteq {D'}$ where ${D'}$ are
named concepts. Table~\ref{tab:tbox_completion} shows the results. We can
see that \method{} consistently outperforms baselines in all metrics. 

\begin{table}[t]
  \centering
  \caption{TBox completion task over axioms $C\sqsubseteq D$ in GO and FoodOn.}~\label{tab:tbox_completion}
  \begin{tabular*}{\columnwidth}{@{\extracolsep\fill}lrrrrrrrr@{\extracolsep\fill}}
    \toprule
    \multirow{2}{*}{Method} & \multicolumn{4}{c}{GO} & \multicolumn{4}{c}{FoodOn} \\ \cmidrule{2-9}
                      & MR         & H@10       & H@100       & AUC        & MR         & H@10       & H@100      & AUC \\
    \midrule
    ELEmbeddings      & \snd{3562} & \snd{0.19} &  \snd{0.37} & \snd{0.92} & 3336       & \snd{0.25} & \snd{0.38} & \snd{0.88} \\
    Box$^2$EL         & 6621       & 0.01       & 0.07        & 0.85       & \fst{2763} & 0.06       & 0.19       & \fst{0.90} \\
    FALCON (5 models) & 8982       & 0.02       & 0.08        & 0.79       & 3815       & 0.02       & 0.12       & 0.86 \\
    \method{}         & \fst{2968} & \fst{0.22} & \fst{0.58}  & \fst{0.93} & \snd{2764} & \fst{0.30} & \fst{0.47} & \fst{0.90} \\
    \bottomrule
  \end{tabular*}
  
\end{table}

\subsection{Protein--Protein Interaction Prediction}

Protein-protein interactions (PPIs) are direct or indirect
interactions between proteins, and information about PPIs is useful in
systems biology and network-based bioinformatics methods. While PPIs
can be investigated experimentally, several strategies have been
developed to predict them using a variety of information, including
the predicted or experimentally determined functions of proteins. The
functions of proteins can be represented using the GO, and if $X$ is a
class from the GO, the axiom $\sub{p_1}{\exists hasFunction. X}$
asserts that the class of proteins $p_1$ has function $X$. PPIs can be
encoded in axioms $interacts(p_1, p_2)$ where $p_1, p_2$ are
proteins. In order to apply our method, we need to ensure that
elements $\exists interacts.p_i$ exists in the lattice for any class
of proteins $p_i$. Therefore, we added the relationships
$\bot \to \exists interacts.p_i$ and $\exists interacts.p_i \to \top$
to the lattice structure for all classes of proteins $p_i$.  We used
the PPI dataset provided in \cite{mowl}.
We compare our method against state-of-the-art methods such as
ELEmbeddings and Box$^2$EL.  \cite{semantic_similarity}, \cite{boxel},
We show the results in Table~\ref{tab:ppi_yeast}, where we can see that
\method{} is not able to outperform over baselines. The PPI benchmark
relies on the assumption that the information GO acts as background
knowledge to predict protein--protein interactions. To further
investigate on this task, we evaluate how well the methods capture the
hierarchy of GO functions, which are axioms of the type
$C \sqsubseteq D$. We compute the deductive closure of axioms
$C\sqsubseteq D$ using the ELK reasoner~\cite{elk}, and evaluate the
capability of each method to generate the axioms in this new set. We
find that ELEmbeddings and Box$^2$EL do not capture the semantics of
GO axioms at all, yet they can perform PPI predictions. Originally,
ELEmbeddings and Box$^2$EL are trained with negative samples just for
PPI axioms, which can cause the other axioms types to converge to a
trivial solution. Since \method{} uses negative samples for all
relationships in the lattice, it can predict PPIs while capturing
other type of information in GO. Our analysis shows that predicting
PPIs on its own is not sufficient to show that a particular embedding
method is utilizing the background knowledge. Further analysis on
embedding methods should be required, which is out of the scope of
this work.

\begin{table*}[t]
  \centering
  \caption[PPI on Yeast.]{Protein-protein interaction prediction on
    Yeast. Left-side shows the results on PPI axioms. Right side shows
    the results on axioms $C \sqsubseteq D$ that are learned during
    training.}\label{tab:ppi_yeast}

  \begin{tabular*}{\linewidth}{@{\extracolsep\fill}lrccccc|ccc@{\extracolsep\fill}}
    \toprule
    \multirow{2}{*}{Method}      & \multicolumn{6}{c|}{PPI axioms $r(a,b)$} & \multicolumn{3}{c}{Axioms $C \sqsubseteq D$} \\ \cmidrule{2-10}
    & MR        & MRR        & H@3        & H@10       & H@100       & AUC        & MR         &  H@100     & AUC \\ \midrule
    ELEmbeddings & 289       & \snd{0.10} & \snd{0.09} & \snd{0.25} & \snd{0.73}  & 0.95       & 23812      &  \snd{0.00}      & 0.53 \\
    Box$^2$EL    & \fst{188} & \fst{0.17} & \fst{0.19} & \fst{0.43} &  \fst{0.81} & \fst{0.97} & \snd{23234}      & \snd{0.00}       & \snd{0.54} \\
    CatE         & \snd{223} & 0.08 & 0.07 & 0.18 & 0.69 & \snd{0.96}       & \fst{8936} & \fst{0.28} & \fst{0.82} \\
    \bottomrule
  \end{tabular*}
  \label{tab:results_ppi}
\end{table*}

\subsection{Effect of partial saturation procedures}

To analyze the impact of the saturation procedures, we extend the
lattices of the ORE1, GO and FoodOn use cases. We first experiment
with the ORE1 lattice as it is the smallest one and apply three types
of saturation: (a) S1, which consists of applying Equations
\ref{eq:24}, \ref{eq:25}, \ref{eq:26}, (b) Tr, which consists on computing the
transitive closure of the lattice, and (c) S1-Tr, which consists on
performing S1 followed by Tr. For GO and FoodOn use cases, which
produce larger lattices, we only apply S1 because the other settings
introduce a large number of elements and morphisms which make the
optmization costly and also hinders the hyperparameter search. We show
performance results in Table~\ref{tab:saturation} and notice that the
S1 procedure contributes to improve the Mean Reciprocal Rank and
Hits@3 metrics in the three use cases. Additionally, for ORE1, the Tr
procedure improves metrics such as Mean Rank and Hits@100; however,
the combination of S1-Tr does not contribute to improve the
performance.

\begin{table}[h]
  \centering
  \caption{Impact of the application saturation procedures on the performance of generation of axioms.}~\label{tab:saturation}
  \begin{tabular*}{\columnwidth}{@{\extracolsep\fill}lrrrrr@{\extracolsep\fill}}
    \toprule
    Method                  & MRR         & H@3         & H@10        & H@100       & MR \\
    \midrule
                            & \multicolumn{5}{c}{ORE1} \\ \cmidrule{2-6}
    \method{}               & \snd{0.175} & 0.097       & \fst{0.505} & \snd{0.958} & 37 \\
    \method{}-S1            & \fst{0.176} & \fst{0.115} & \snd{0.426} & 0.884       & 46 \\
    \method{}-Tr            & 0.164       & \snd{0.104} & 0.381       & \fst{0.991} & \fst{23} \\
    \method{}-S1-Tr         & 0.155       & 0.098       & 0.367       & 0.931       & \snd{30} \\
    \midrule
                            & \multicolumn{5}{c}{GO} \\ \cmidrule{2-6}
    \method{}               & \snd{0.062} & \snd{0.008} & \snd{0.216} & \snd{0.578} & \fst{2968} \\
    \method{}-S1            & \fst{0.066} & \fst{0.011} & \fst{0.226} & \fst{0.595} & \snd{3002} \\
    \midrule
    \multirow{2}{*}{Method} & \multicolumn{5}{c}{FoodOn} \\
    \cmidrule{2-6}
    \method{}               & \snd{0.087} & \snd{0.023} & \fst{0.298} & \fst{0.473} & \fst{2764} \\
    \method{}-S1            & \fst{0.094} & \fst{0.121} & \snd{0.238} & \snd{0.419} & \snd{3310} \\
    \bottomrule
  \end{tabular*} 
\end{table}

\subsection{Effect of hyperparameters}

\begin{figure}[h]
  \includegraphics[width=\textwidth]{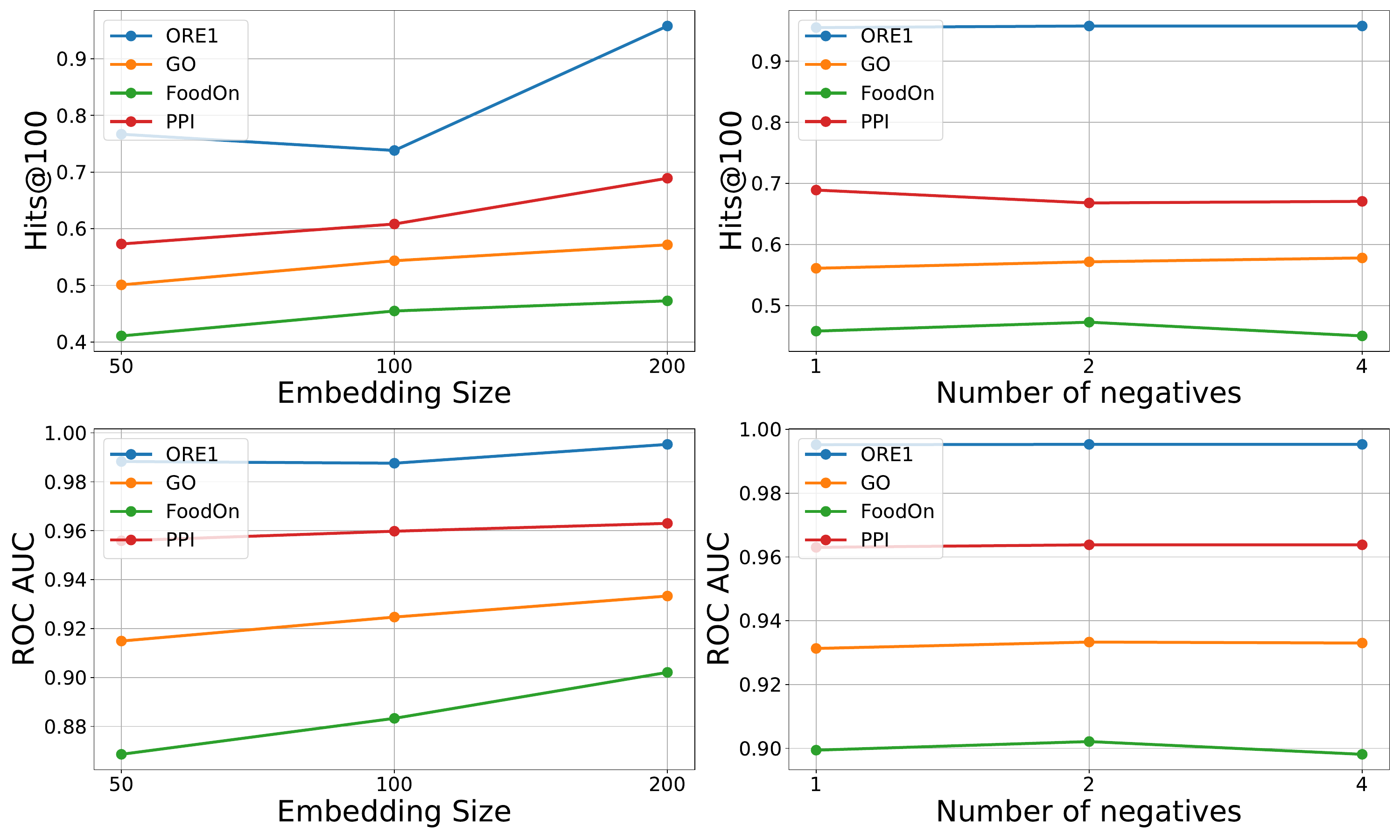}
  \caption{Impact of embedding size and number of negatives on the
    Hits@100 and ROC AUC over different datasets.} \label{fig:hyp}
\end{figure}

The time and space complexity of \method{} increases linearly with the
number of operators. However, the number of operators can be
arbitrarily large for axioms in $\mathcal{ALC}$. Furthermore,
hyperparameters such as embedding size and number of negative samples
can have an impact on training and/or inference time as well as on
memory consumption. In Table~\ref{fig:hyp}, we analyze how these
hyperparameters impact on performance. We chose Hits$@$100 and ROC AUC
metrics and show that while the embedding dimension has a direct
impact performance (the higher the dimension the better the
performance), the number of negative samples does not have large
effect (either positive or negative).

\section{Discussion}

We have developed a method named CatE that generates embeddings for the
$\mathcal{ALC}$ language. CatE consists on materializing the lattice structure
of concept descriptions found in a $\mathcal{ALC}$ knowledge
base. Furthermore, we use an order-preserving loss function to optimize
the embedding space, and we show that when our loss function is
minimized, the embedding space preserves partial order properties. We
have applied our method to different forms of knowledge base
completion tasks, and we showed that our method can outperform several
state-of-the-art methods.

Additionally, we implemented saturation procedures to extend the
lattices and the information therein. We showed that saturated
versions of the lattices can improve on some metrics. However, not all
the saturation rules can be applied if the knowledge bases are large
because the size of the resulting lattice and the number of morphisms
can hinder the application of the optimization process. A potential
direction for future work can be to generating some concepts directly
in the embedding space rather than explicitly materializing them
within the lattice.

Current graph-based methods to embed DL knowledge bases (ontologies)
construct graphs relying on syntactic information therein and the
embedding process is not guaranteed to be invertible. On the other
hand, methods such as ELEmbeddings, Box$^2$EL and FALCON are able to
generate approximate models for DL knowledge bases. We state that
\method{} stands in a midpoint between both types of methods. \method{}
looks into the syntactical information in the knowledge base to
construct a lattice and, consequently, an embedding space that is
consistent to the semantics.

However, as in graph-based embeddings, \method{} cannot make
inferences over concepts that are not explicitly stated in the
lattice. This is a limitation that was exposed in the protein--protein
interaction task, where we had to add concept descriptions a priori in
order to be able to make inferences over them. To mitigate this issue,
future work can focus on solutions based on inductive learning over
knowledge graphs, which can be applicable in the context of lattices.

\section{Conclusion}

We developed an embedding method for the  $\mathcal{ALC}$ that
preserves the lattice structure of concept descriptions. Our method
materializes the lattice structure following connections between
Description Logics and Category Theory. The lattice in place is
embedded into an ordered-vector space. We provide empirical results
that our method can perform effectively across different tasks
involving knowledge base completion.

\section{Acknowledgement}

We acknowledge support from the KAUST Supercomputing Laboratory.

\section{Funding}

This work has been supported by funding from King Abdullah University
of Science and Technology (KAUST) Office of Sponsored Research (OSR)
under Award No. URF/1/4355-01-01, URF/1/4675-01-01, URF/1/4697-01-01,
URF/1/5041-01-01, and REI/1/5334-01-01.  This work was supported by
the SDAIA--KAUST Center of Excellence in Data Science and Artificial
Intelligence (SDAIA--KAUST AI), by funding from King Abdullah
University of Science and Technology (KAUST) -- KAUST Center of
Excellence for Smart Health (KCSH) under award number 5932, and by
funding from King Abdullah University of Science and Technology
(KAUST) -- KAUST Center of Excellence for Generative AI under award
number 5940. 

%%%%%%%%%%% The bibliography starts:

%%  ios1.bst will be used to                               %%

\nocite{*}
% if your bibliography is in bibtex format, use those commands:
% \bibliographystyle{ios1}           % Style BST file.
% \bibliography{refs}        % Bibliography file (usually '*.bib')
\printbibliography
% or include bibliography directly:
%\begin{thebibliography}{0}
%\bibitem{r1} F. Author, Information about cited object.
%
%\bibitem{r2} S. Author and T. Author, Information about cited object.
%\end{thebibliography}

\clearpage
\appendix

\section{Hyperparameter Selection}~\label{app:hyperparameter}

\begin{table}[h]
  \centering
  \caption{Selection of hyperparameters for the different methods with respect to the dataset used. E.S.: Embedding size, L.R.: learning rate, M: margin, B.S.: batch size, N.N.: number of negative samples.} 
  \begin{tabular*}{\columnwidth}{@{\extracolsep\fill}lrrrrr@{\extracolsep\fill}}
      \toprule
    {Method} & \centered{E.S.} & \centered{L.R.} & \centered{M} & \centered{B.S.} & {N.N.} \\
      \midrule
                            & \multicolumn{5}{c}{GO}\\
    \midrule
    ELEmbeddings & 200 & 0.0001 & 0.1 & 20000 & 1\\
    Box$^2$EL & 200 & 0.00001 & 0.1 & 20000 & 1\\
    CatE & 200 & 0.00001 & 1 & 32768 & 4\\
    \midrule
                            & \multicolumn{5}{c}{FoodOn} \\ \midrule
    ELEmbeddings & 50 & 0.001 & 0.1 & 20000 & 1\\
    Box$^2$EL & 200 & 0.0001 & 0.1 & 40000 & 1\\
    CatE & 200 & 0.0001 & 1 & 8192 & 2\\ \midrule
                            & \multicolumn{5}{c}{ORE1} \\ \midrule

    ELEmbeddings & 200 & 0.00001 & 0.01 & 4096 & 1\\
    Box$^2$EL & 200 & 0.0001 & 0 & 8192 & 1\\
    CatE & 200 & 0.0001 & 1 & 32768 & 4\\\midrule
             & \multicolumn{5}{c}{PPI} \\\midrule
    CatE & 256 & 0.0001 & 0.1 & 2048 & 4 \\
    \bottomrule
  \end{tabular*}
  \label{tab:unsat}
\end{table}

\end{document}